\documentclass{ceurart}

\usepackage{listings}
\usepackage{graphicx}
\usepackage{subcaption}

\begin{document}

\copyrightyear{2026}
\copyrightclause{Copyright for this paper by its authors.
	Use permitted under Creative Commons License Attribution 4.0
	International (CC BY 4.0).}

\conference{SuRE'26: Workshop on Sustainability and Resource-Efficiency of Artificial Intelligence, August 17, 2024, Bremen, Germany}

\title{WattGPU: Predicting Inference Power and Latency on Unseen GPUs and LLMs}

\author[1]{Mauricio {Fadel Argerich}}[%
	orcid=0009-0008-9348-8426,
	email=mauricio.fadel@alumnos.upm.es,
]
\cormark[1]
\address[1]{Universidad Politécnica de Madrid, C. de los Ciruelos, 28660 Boadilla del Monte, Madrid, Spain
	}

\author[2]{Jonathan F\"urst}[%
	orcid=0000-0002-4062-1827,
	email=jonathan.fuerst@zhaw.ch,
]
\address[2]{Zurich University of Applied Sciences, Gertrudstrasse 15, 8400 Winterthur, Switzerland}

\author[1]{Marta Pati\~no-Mart\'inez}[%
	orcid=0000-0003-2997-3722,
	email=mpatino@fi.upm.es,
]

\cortext[1]{Corresponding author.}

\begin{abstract}
Large Language Model (LLM) inference workloads are a rapidly growing contributor to data center energy consumption. Optimizing these deployments requires matching specific LLMs to the most efficient GPUs, but operators currently lack the tools to do so without exhaustively profiling each combination. While some predictive models exist, they still require profiling data and struggle to generalize to hardware unseen during training. To address this, we introduce \textit{WattGPU}, featuring two predictive models for mean GPU power draw and Inter-Token Latency (ITL). Our approach leverages only publicly available LLM metadata and GPU specifications, eliminating the need for hardware access or profiling while enabling generalization to unseen NVIDIA server-grade GPUs and LLMs. We evaluate our models using rigorous leave-one-GPU-out and leave-one-LLM-out cross-validation on a dataset of 42 open-source LLMs (0.1B--27B parameters) and 8 GPUs under both offline and server scenarios. The mean power draw model achieves a median absolute percentage error of $\leq3.4\%$ for offline and $\leq13.5\%$ for server scenarios on unseen GPUs, while the latency model achieves $\leq8.5\%$ in server mode, both maintaining strong GPU ranking correlations for server scenarios (Kendall $\tau\geq0.76$). Compared to standard physically grounded baselines ---Load-Scaled Thermal Design Power (TDP) for power draw and roofline for latency--- our models reduce median absolute percentage error by approximately 4$\times$ on unseen LLM-GPU combinations for server scenarios or approximately 2$\times$ for completely unseen GPUs. WattGPU's data and code are publicly available at  \url{https://github.com/maufadel/wattgpu}.
\end{abstract}


\maketitle

\section{Introduction}
The growing use of Large Language Models (LLMs) in user and enterprise applications, combined with the increasing number and quality of open-source models, is driving the proliferation of LLM inference servers in cloud deployments. This trend is further reinforced by user and business demand for data privacy and sovereignty. 

While attention typically focuses on hyperscalers and major AI companies (e.g., OpenAI, AWS, Google) that consume gigawatts of energy to process billions of daily requests, small- to medium-scale deployments are rapidly growing. These smaller deployments are often less optimized, and their energy consumption remains poorly characterized. Consequently, even though their individual energy usage is lower, their aggregate energy footprint is growing alarmingly fast~\cite{iea2025energyai}. This inefficiency is particularly pronounced in serving workloads; recent measurements reveal that even within the online deployments of industry leaders, GPUs spend 14--76\% of their time and 7--65\% of their energy in execution-idle states~\cite{lei2026energy}. This hints at massive potential savings if deployment efficiency can be improved.

\begin{figure}[htbp]
  \centering
\includegraphics[width=0.5\linewidth]{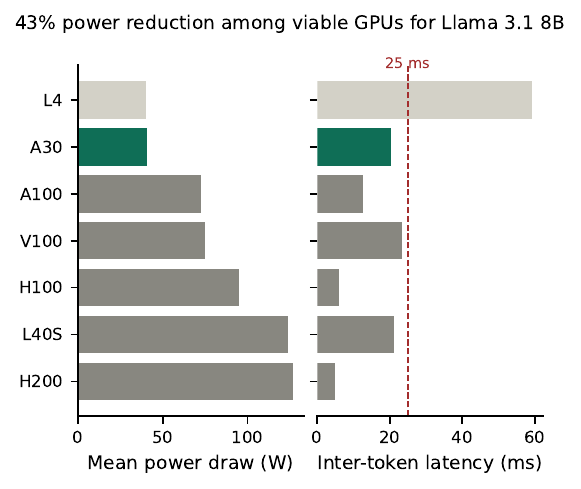}
  \caption{GPU selection is a key lever for energy savings in LLM inference. For deploying Llama 3.1 8B with a Service Level Agreement (SLA) of a mean ITL $<25$ms, it is possible to reduce power draw by 43\% by choosing an A30 instead of an H100. Data from the Watt Counts dataset showing only GPUs where the LLM fits in memory.}
  \label{fig:gpu-spread}
\end{figure}

A critical lever for optimizing energy consumption in LLM inference is optimal GPU selection. GPUs dominate the energy consumption of LLM inference~\cite{theodorou2024energy}, and careful LLM-GPU matching yields substantial efficiency gains. For instance, deploying a medium-sized model like Llama 3.1 8B on an NVIDIA A30 can reduce power draw by up to 43\% in low-load scenarios compared to an H100 (a common high-end GPU), as illustrated in Figure~\ref{fig:gpu-spread}. Achieving these savings not only reduces operational costs but also mitigates the growing carbon footprint of LLM serving, which currently raises energy prices~\cite{burian2025_increasing_ai_energy} and hinders progress toward the Paris Agreement and UN Agenda 2030 targets.

However, identifying these optimal deployment configurations currently requires exhaustive manual profiling of multiple candidate LLM-GPU combinations. This profiling process is prohibitively expensive: it demands physical access to diverse hardware, specialized expertise, and significant economic resources---luxuries that most small- and medium-scale operators lack. 

To close this gap, we propose \textit{WattGPU}, a method for predicting the inference power draw and latency of LLM-GPU pairs that once trained, requires zero hardware access or profiling, leveraging only publicly available metadata. Our main contributions are:

\begin{enumerate}
    \item \textbf{To the best of our knowledge, the first predictive models for LLM inference power draw and latency evaluated on unseen GPUs.} We introduce two predictive models for LLM inference characterization: \textbf{mean GPU power draw} and \textbf{Inter-Token Latency (ITL)}. Both models rely exclusively on publicly available LLM metadata and GPU manufacturer specifications. These models, alongside their training and evaluation pipelines, are open-sourced in the accompanying repository.\footnote{\url{https://github.com/maufadel/wattgpu}}

    \item \textbf{Rigorous generalization evaluation.} We evaluate the models' capacity to generalize to unseen hardware and models across 8 server-grade NVIDIA GPUs and 42 LLMs (0.1B--27B parameters) using Leave-One-GPU-Out and Leave-One-LLM-Out cross-validation. Our models achieve median absolute percentage errors on unseen GPUs of $\leq3.4\%$ for offline and $\leq13.5\%$ for server scenarios for mean power draw. Our latency model achieves $\leq8.5\%$ error in server mode, and both models achieve a Kendall $\tau\geq0.76$ for GPU rankings for server scenarios.

    \item \textbf{Demonstrated improvement over analytical baselines.} We compare our approach against physically motivated baselines (two TDP-derived baselines for power, roofline for ITL). We demonstrate that our learned models successfully capture complex overheads beyond first-order analytical effects, effectively improving rankings for GPUs across both scenarios in cross validation evaluation ---for unseen (GPU, LLM) combinations--- and reducing the median absolute percentage error by approximately 4$\times$ on unseen LLM-GPU combinations or approximately 2$\times$ for completely unseen GPUs compared to the strongest baselines for server scenarios.
\end{enumerate}

The remainder of this paper is organized as follows. Section~\ref{sec:related} analyzes related work and its limitations. Section~\ref{sec:background} introduces the metrics, the inference scenarios, and the dataset used throughout this work. Section~\ref{sec:method} presents the two predictive models and describes their features. The evaluation of these models is presented in Section~\ref{sec:evaluation}. Section~\ref{sec:discussion} discusses the implications, current challenges, and promising future directions for our work, and finally, Section~\ref{sec:conclusion} summarizes our key findings.

\section{Related Work}
\label{sec:related}
We review prior work on predicting and measuring LLM inference performance across two categories: energy measurement and benchmarking, and predictive modeling of energy and latency. In both cases, existing approaches require profiling the target LLM-GPU combination, limiting their utility for operators selecting among deployment options they have not yet profiled, and their ability to generalize to unseen GPUs has not been evaluated.

\paragraph{Energy Measurement and Benchmarking.} Numerous studies have profiled the energy and carbon costs of LLMs during training and inference \cite{strubell2020energy,samsi2023words,argerich2024measuring}. Systematic efforts like MLPerf Power \cite{tschand2025mlperf} and leaderboards such as AI Energy Score~\cite{aienergyscore-leaderboard} and ML.ENERGY~\cite{mlenergy-neuripsdb25}, provide empirical data but rely on physical power meters or extensive profiling and provide data only for two high-end GPUs (NVIDIA H100 and NVIDIA B200). Recently, TokenPowerBench~\cite{niu2026tokenpowerbench} introduced a lightweight benchmark that reduces the profiling needed for estimating energy consumption of LLM inference, but it still requires profiling and hardware access. Further, these approaches cannot estimate energy for hardware or models not profiled. From a complementary angle, Krupp et al.~\cite{krupp2026taking} use inference time as a proxy to infer the utilized GPU and estimate energy for LLM APIs, where hardware is often unknown. In contrast, our work predicts both token latency and power draw from GPU specifications and LLM metadata to inform deployment decisions.

\paragraph{Predictive Modeling of Energy and Latency.} Fu et al.~\cite{fu2025llmco2} use a graph neural network and Roofline-based GPU features to model the energy and carbon footprint of LLM inference and evaluate their approach with three LLMs and two GPUs. Wilkins et al.~\cite{wilkins2024offline} develop a workload-based energy model for LLMs on a multi-GPU NVIDIA A100 system. throttLL'eM~\cite{kakolyris2025throttll} optimizes cluster energy by predicting usage based on profiling data gathered for specific hardware (e.g., H100 or A100 clusters). Other analytical models (e.g., Patwari et al.~\cite{patwari2025forecasting} for latency) and simulators (e.g., Vidur~\cite{agrawal2024vidur} for latency and its extension~\cite{ozcan2025quantifying} for energy) require profiling-derived values or training data, offering limited generalization across different engine configurations or LLMs. Imai et al.~\cite{imai2024predicting} combine regression with an analytical model to estimate roofline token latency prediction on a single-GPU for vLLM and Triton. However, the approach is evaluated on a single GPU (NVIDIA A100), leaving generalization to unseen GPUs as an open question. EcoLogits~\cite{rince2025ecologits} estimates the energy and CO\textsubscript{2} emissions of LLMs based on a fitted formula to ML.ENERGY data; however, its estimations are for steady-state server measurements, ignoring idle and transient power, and are limited to NVIDIA H100 GPUs only, as it cannot generalize to unseen hardware. The closest prior work is by Caravaca et al.~\cite{caravaca2025prompts}, who use ML to predict energy per prompt, but their validation is limited to a single unseen LLM and only known GPUs. 
\section{Background}
\label{sec:background}
This section introduces the operational scenarios for LLM inference, the metrics used to characterize energy and latency, and the dataset utilized for training and evaluation.

\subsection{Inference Scenarios}
We analyze LLM inference under two operational scenarios, modeled after the MLPerf Inference benchmark suite~\cite{tschand2025mlperf}: offline, and server. In offline scenarios, a large number of requests is typically submitted concurrently to the system (i.e., a batch of requests), and the system processes them at maximum throughput, maximizing GPU utilization. In server scenarios, user requests arrive according to a load pattern (e.g., a Poisson distribution); the system is continuously waiting for user requests to arrive and needs to process them under certain latency requirements. In this case, the request load significantly impacts the GPU utilization rate and energy per token.

\subsection{Energy Metrics}
Energy consumption for LLM inference depends jointly on power draw and generation latency. We characterize energy consumption for LLM inference using two relevant metrics: \textbf{Mean power draw $\bar{P}$} (in Watts), defined as the average GPU power draw sustained during inference, and \textbf{Inter-Token Latency (ITL)} (in seconds), the average time between the generation of consecutive tokens. Throughout this paper, $\bar{P}$ refers to mean GPU power draw only; we focus on GPUs as prior work has indicated them as the dominant power consumer in LLM inference~\cite{argerich2024measuring,theodorou2024energy}.

The calculation of ITL depends on the deployment scenario. In the server scenario, ITL excludes Time-to-First-Token (TTFT), representing latency in the decode phase. It is calculated as: 
\begin{equation}
    ITL=\frac{L-TTFT}{R_{tok}-1}
\end{equation}
where $L$ is end-to-end latency from request arrival until the last output token is generated and $R_{tok}$ is the number of generated tokens for the response. ITL is computed per request and averaged across all completed requests in the experiment. In the offline scenario, where TTFT is not available, ITL is calculated as total wall time divided by the number of tokens generated, so prefill is amortized across all tokens:
\begin{equation}
    ITL=\frac{W}{N_{tok}}
\end{equation}
where $W$ is wall time and $N_{tok}$ is the total number of generated tokens.

Metrics per token are preferred over totals because they are independent of prompt and response length, enabling the comparison of the performance of LLM-GPU pairs for heterogeneous workloads. In the offline scenario, it is possible to derive energy per token using the output of both models presented here: $E_{tok}=\bar{P} \cdot ITL$ in Joules. In server scenarios, this product is not a meaningful efficiency metric as $\bar{P}$ averages over idle and busy intervals, while ITL characterizes per-token latency only during active generation. We therefore frame server scenario efficiency as a multi-objective problem where $\bar{P}$ characterizes energy consumption over time and ITL characterizes user experience. 
\section{Method}
\label{sec:method}
We implement two regression models to predict mean power draw $\bar{P}$ and Inter-Token Latency (ITL) using features derived entirely from publicly available data. By completely avoiding runtime metrics, our approach enables pre-deployment estimation without dedicated profiling runs.

\subsection{Feature Engineering and Selection}
To characterize GPU-LLM pairs, we extract hardware specifications from manufacturer-reported data (e.g., Thermal Design Power, memory bandwidth, transistor count) and LLM architectural details from Hugging Face metadata (e.g., parameter count, number of layers, attention heads). We initially considered a broad set of features recommended by existing literature~\cite{argerich2026watt} and empirically refined them by evaluating Leave-One-GPU-Out (LOGO) and Leave-One-LLM-Out (LOLO) performance. The final selected features for both models are summarized in Table~\ref{tab:features}.

To better capture real-world hardware behavior and power draw dynamics, we also engineered three derived features:
\begin{itemize}
    \item \textbf{Boosting ratio:} Encodes frequency scaling by quantifying the ratio between the GPU's boosted clock and base clock. This scaling affects power draw and serves as a baseline to measure expected power transitions between idle and high utilization states.
    \item \textbf{Bandwidth Latency:} Represents the time required to read all model weights from memory once. It is calculated as the model size in GB (estimated as the total number of parameters $\times$ 2 bytes for FP16) divided by the GPU's memory bandwidth.
    \item \textbf{Compute Latency ($C_{\mathrm{lat}}$):} The theoretical lower bound on the time needed to compute each token, assuming full FLOPS utilization. It is calculated as:
    \begin{equation}
        C_{\mathrm{lat}}=\frac{2P_{\mathrm{model}}\cdot 10^{9}}{FLOPS_{GPU}}
    \end{equation}
    where $FLOPS_{GPU}$ is the maximum FP16 FLOPS of the GPU, and $P_{\mathrm{model}}$ is the number of parameters in the model in billions.
\end{itemize}

\begin{table}[htbp]
\centering
\caption{Features used in the predictive models for Mean Power Draw ($\bar{P}$) and Inter-Token Latency (ITL). All features are derived or directly taken from public specifications without any runtime profiling.}
\label{tab:features}
\scriptsize
\begin{tabular}{llccc}
\toprule
\textbf{Category} & \textbf{Feature} & \textbf{Data Source} & \textbf{Power $\bar{P}$} & \textbf{ITL} \\
\midrule
\textbf{GPU} & Base Clock Frequency & GPU Specs & \checkmark & \\
 & Boosting Ratio & Derived & \checkmark & \\
 & Transistor Count & GPU Specs & \checkmark & \\
 & Memory Bandwidth & GPU Specs & \checkmark & \checkmark \\
 & Release Year & GPU Specs & \checkmark & \checkmark \\
 & Memory Size & GPU Specs & & \checkmark \\
 & FP16 (T)FLOPS & GPU Specs & & \checkmark \\
 & Memory Type & GPU Specs & & \checkmark \\
\midrule
\textbf{LLM} & Model Type & HF Metadata & \checkmark & \checkmark \\
 & Parameters (log) & Derived & \checkmark & \checkmark \\
 & Number of Layers & HF Metadata & \checkmark & \checkmark \\
 & KV Heads & HF Metadata & \checkmark & \checkmark \\
 & Attention Heads & HF Metadata & & \checkmark \\
 & Hidden Size & HF Metadata & & \checkmark \\
\midrule
\textbf{Context} & Operational Scenario & Setup & \checkmark & \checkmark \\
\textbf{\& Derived} & Bandwidth Latency & Derived & \checkmark & \checkmark \\
 & Compute Latency & Derived & & \checkmark \\
\bottomrule
\end{tabular}%
\end{table}


\subsection{Model Selection and Training}
To facilitate generalization across heterogeneous GPU architectures, we normalize the target variable for the power model (GPU power draw) by dividing it by the GPU's Thermal Design Power (TDP) or the GPU's configured power limit. The TDP measures the nominal thermal dissipation target of the GPU and is strongly correlated with its long-term maximum power draw. Predicting this relative metric allows the model to learn scale-invariant power behavior, directly enabling generalizability to unseen GPUs.

For predicting both $\bar{P}$ and ITL, we evaluated several algorithms, including linear regression, elastic nets, multi-layer perceptrons, and gradient-boosted regression. Gradient-boosted regression trees (XGBoost) consistently achieved the best performance in Median Absolute Percentage Error (MdAPE) and Pearson $r$ during LOGO cross-validation. 

Consequently, both models are implemented using XGBoost with \texttt{reg\_lambda=100}. For the power draw model, the maximum tree depth was set to \texttt{max\_depth=6} and number of estimators to \texttt{n\_estimators=100}, while for the ITL model \texttt{max\_depth=5} and \texttt{n\_estimators=200} were used. These hyperparameters were selected via grid search over \texttt{max\_depth} $\in \{3, 4, 5, 6, 7\}$, \texttt{reg\_lambda} $\in \{0.1, 1, 10, 50, 100, 500\}$, \texttt{n\_estimators} $\in \{50, 100, 200, 500\}$, with all other hyperparameters left at their default values.

\section{Experimental Evaluation}
\label{sec:evaluation}

\begin{table*}[ht]
    \centering
    \caption{NVIDIA GPUs used in the dataset and their main technical characteristics: architecture, profile (consumer or enterprise), memory, TDPs (or their configured power limits), tensor FP16 TFLOPS without sparsity, memory bandwidth, L2 cache size, TFLOPS (without sparsity) per Watt, and release year.}
    \label{tab:gpus}
    \resizebox{0.8\linewidth}{!}{%
    \begin{tabular}{llccccccc}
    \toprule
    GPU & Architecture & Memory (GB) & TDP (W) & TFLOPS (FP16) & Mem BW (GB/s) & L2 Cache (MB)& TFLOPS/Watt & Year \\
    \midrule
    Tesla V100 SXM2 & Volta & 32 & 250 & 125 & 898 & 6 & 0.5 & 2018 \\
    Tesla T4 & Turing &  16 & 70 & 65 & 320 & 4 & 0.93 & 2018 \\
    A100 SXM4 & Ampere &  40 & 400 & 312 & 1555 & 40 & 0.78 & 2020 \\
    A30 PCIe & Ampere &  24 & 165 & 165 & 933 & 24 & 1.00 & 2021 \\
    L40S & Ada Lovelace &  48 & 350 & 362 & 864 & 48 & 1.03 & 2022 \\
    L4 & Ada Lovelace &  24 & 72 & 121 & 300 & 48 & 1.68 & 2023 \\
    H100 NVL & Hopper &  94 & 400 & 835 & 3940 & 50 & 2.09 & 2023 \\
    H200 NVL & Hopper &  141 & 700 & 835 & 4800 & 50 & 1.19 & 2024 \\
    \bottomrule
    \end{tabular}
    }
\end{table*} 

We assess the accuracy and generalization capabilities of both models for Mean power draw $\bar{P}$ and Inter-Token Latency (ITL) for offline, low-, and high-load server scenarios using three strategies of increasing difficulty:
\begin{enumerate}
    \item \textbf{5-fold Cross Validation (CV)}: the dataset is partitioned into 5 folds and each time, 4 are used for training and the remaining one for test. Since the dataset includes multiple repetitions for each (GPU, LLM) combination, we use GroupKFold keyed on these columns to prevent replicate leakage between train and test folds. In this case, each GPU and each LLM has been seen individually in training but not in that specific combination. \emph{This evaluates generalization to unseen (GPU, LLM) combinations  of known GPUs and LLMs}.
    \item \textbf{Leave-One-GPU-Out (LOGO)}: each GPU is left out of the training set in turn and used for testing. \emph{This evaluates generalization to completely unseen GPUs.}
    \item \textbf{Leave-One-LLM-Out (LOLO)}: each LLM is left out of the training set in turn and used for testing. \emph{This evaluates generalization to completely unseen LLMs.}
\end{enumerate}

For each strategy, we report \textbf{Median Absolute Percentage Error (MdAPE)}, robust to outliers from very small target values to assess prediction accuracy in absolute terms; \textbf{Pearson correlation $r$}, computed between predictions and ground truth across all test points to measure linear agreement; and \textbf{Kendall $\tau$} to assess GPU and LLM rankings quality. GPU $\tau$ evaluates how well the model ranks GPUs: for each (LLM, scenario) pair in the test set, we compute the rank correlation between predicted and actual values across GPUs, and report the mean across pairs. LLM $\tau$ evaluates how well the model ranks LLMs, averaging over (GPU, scenario) pairs. Both are computed identically across CV, LOGO, and LOLO; what differs is the composition of the test set in each split.
Higher $\tau$ values indicate better ranking quality regardless of absolute prediction error.

\subsection{The Watt Counts Dataset}
We train and evaluate the predictive models for both $\bar{P}$ and ITL on the Watt Counts dataset~\cite{argerich2026watt}, a public dataset of LLM inference power and latency measurements covering 50 LLMs across 10 NVIDIA GPUs. Watt Counts records $\bar{P}$ under both offline and server scenarios; in the server scenarios, requests follow a Poisson distribution at two different request rates ($\lambda \in \{0.017, 0.33\}$) for low-load and high-load servers respectively.

Because our focus is on cloud scenarios, we exclude 2 consumer-grade GPUs (NVIDIA RTX GeForce 4090 and 3090) in Watt Counts and leave 8 Mixture-of-Experts models as future work, as their inference exhibits different patterns that require separate modeling. The resulting dataset covers 8 different server-grade GPUs shown in Table~\ref{tab:gpus} and 42 dense LLMs from 125M to 27B parameters listed in Table~\ref{tab:llms}.

\subsection{Mean Power Draw Prediction}
We evaluate the predictive model for $\bar{P}$ against two baselines:
\begin{itemize}
    \item \textbf{Plain TDP}: assumes the mean power draw will match the TDP during execution. While simple, this is a common baseline used for comparing the power draw of devices.
    \item \textbf{Load-Scaled TDP}: this baseline estimates power by linearly scaling the GPU's maximum TDP, proportional to a scenario-specific load factor $\rho$:
    \begin{equation}
        \bar{P}_{\mathrm{baseline}} = \rho \cdot TDP
    \end{equation}
    where $\rho = 1.0$ for the offline scenario (representing 100\% utilization), and $\rho \in \{0.2,0.6\}$ for low and high server scenarios respectively. For each scenario we sweep $\rho$ in 0.1 increments and report the value that minimizes the baseline MdAPE, evaluating the baseline under its best-case calibration. This represents a more rigorous baseline that inherently accounts for the operational load distribution.
\end{itemize}

Both baselines are not trained on any data, so the evaluation strategies do not change their results. In addition, they are based on GPU specifications only, which means they cannot rank LLMs. Results are shown in Table~\ref{tab:power_results}.

\begin{table}[h]
\centering
\caption{Mean power draw prediction performance for XGBoost (XGB) under three data-split strategies (CV, LOGO, LOLO), against Load-Scaled TDP (LS TDP) and plain TDP baselines. Both baselines are analytical and depend only on GPU specifications, so their evaluation is invariant to the data split and they cannot rank LLMs (LLM $\tau$ is undefined).}
\label{tab:power_results}
\scriptsize
\begin{tabular}{ccccccc}
\toprule
Model & Scenario & Strategy & MdAPE & Pearson $r$ & GPU $\tau$ & LLM $\tau$ \\
\midrule
XGB & Offline & CV & 1.6\% & 0.992 & 0.96 & 0.44 \\
XGB & Offline & LOGO & 3.4\% & 0.988 & 0.95 & 0.40 \\
XGB & Offline & LOLO & 2.0\% & 0.979 & 0.97 & 0.33 \\
TDP & Offline & — & 4.4\% & 0.916 & 0.96 & — \\
LS TDP & Offline & — & 4.4\% & 0.916 & 0.96 & — \\
\midrule
XGB & Server & CV & 5.5\% & 0.981 & 0.86 & 0.68 \\
XGB & Server & LOGO & 13.5\% & 0.965 & 0.76 & 0.72 \\
XGB & Server & LOLO & 6.7\% & 0.955 & 0.85 & 0.54 \\
TDP & Server & — & 190.1\% & 0.590 & 0.60 & — \\
LS TDP & Server & — & 26.0\% & 0.779 & 0.60 & — \\
\bottomrule
\end{tabular}
\end{table}

The TDP baselines achieve a very good score in the offline scenario (MdAPE=4.4\%) because the power draw tends to match the TDP for extended periods of high utilization --note that in this case, both baselines achieve the same result because for the load-scaled TDP we use $\rho=1$ for the offline scenario. For server scenarios, this changes dramatically: the baseline exhibits an MdAPE of 190.1\%, and the load-scaled TDP 26.0\%, with a GPU $\tau=0.60$, indicating TDP-based heuristics are a limited approach to rank GPUs for server scenarios. Our approach improves greatly on this baseline, achieving an MdAPE of 1.6\% and 5.5\% in CV for offline and server scenarios respectively, with Pearson $r\geq0.98$ and GPU Kendall $\tau\geq0.86$. 

\begin{figure}[ht]
    \centering
\includegraphics[width=0.8\linewidth]{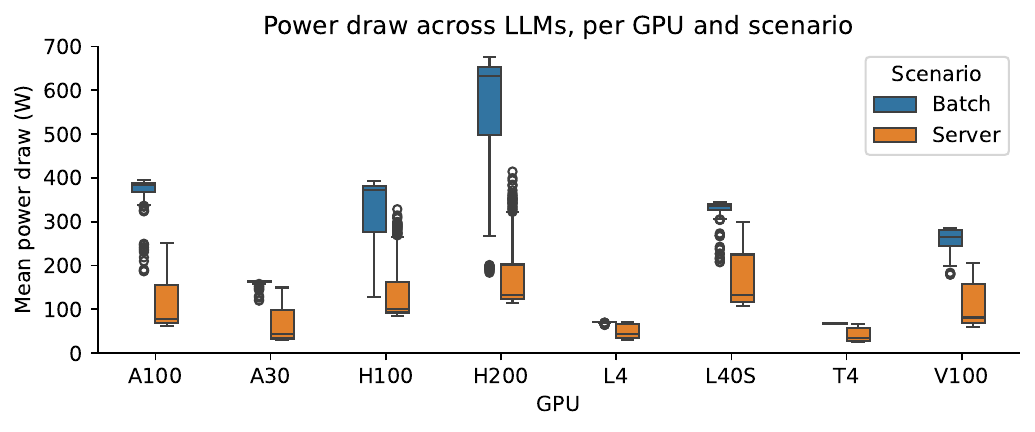}
\caption{Power draw distribution across the 42 LLMs, per GPU and scenario. Each box represents the distribution of the per-LLM mean power draw samples averaged over iterations with the same (GPU, LLM, scenario) configuration. Error bars show $\pm1$ standard deviation across samples.}
\label{fig:power-draw-distribution}
\end{figure}

Ranking LLMs based on their mean power draw is a challenging task, especially in offline mode, as vLLM runs inference with GPU utilization close to 100\% for most GPUs and models, so its power draw tends to the GPU's TDP regardless of which LLM is running, as shown in Figure~\ref{fig:power-draw-distribution}. Notable exceptions are powerful GPUs such as the H200 and the H100, where even vLLM struggles to achieve full utilization when running very small models such as GPT-2 small. In this case, small absolute errors translate to large changes in rankings. Our predictive model still achieves meaningful results even for this case, with LLM $\tau=0.33$, whereas random is $\tau=0$. Note that in this case, the TDP-based baselines cannot be evaluated since they are invariant to LLMs.

\begin{figure}[ht]
    \centering
    \begin{subfigure}{0.48\textwidth}
        \centering
\includegraphics[width=\linewidth]{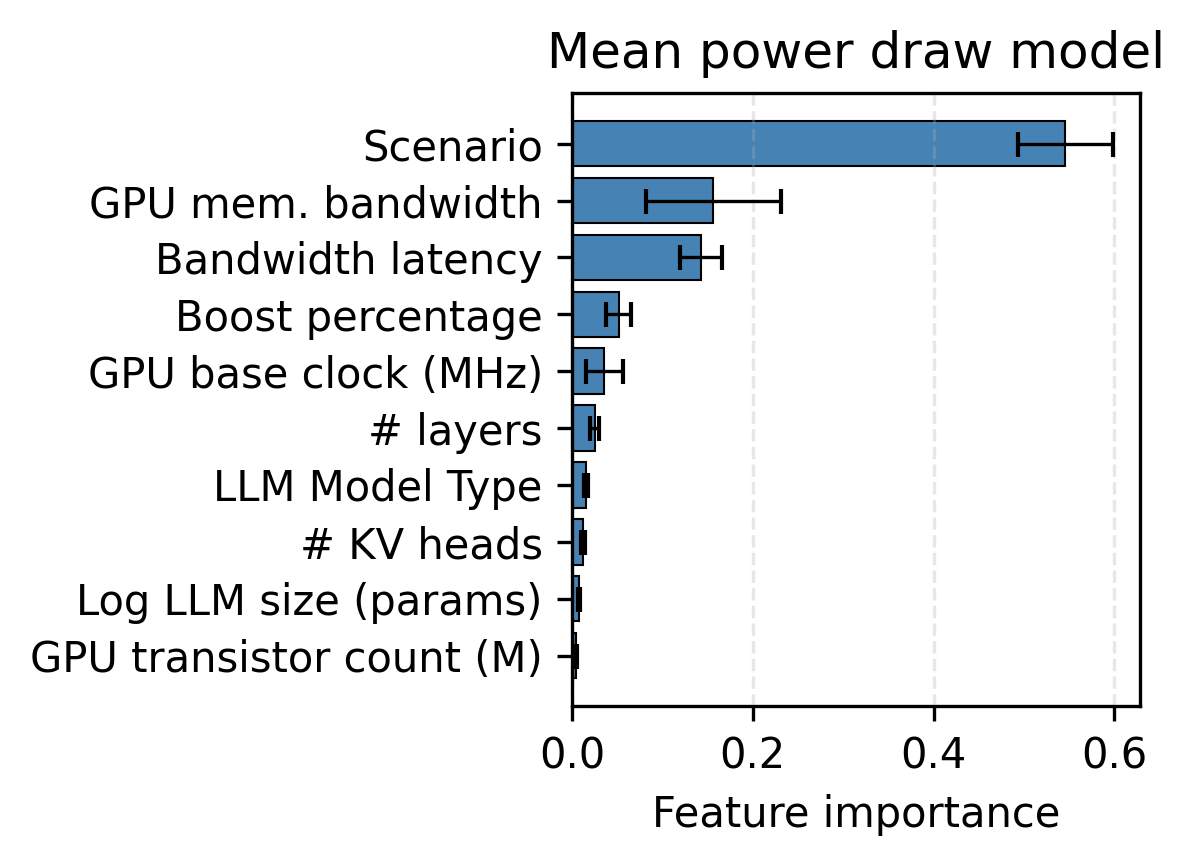}
    \end{subfigure}
    \hfill
    \begin{subfigure}{0.48\textwidth}
            \centering
\includegraphics[width=\linewidth]{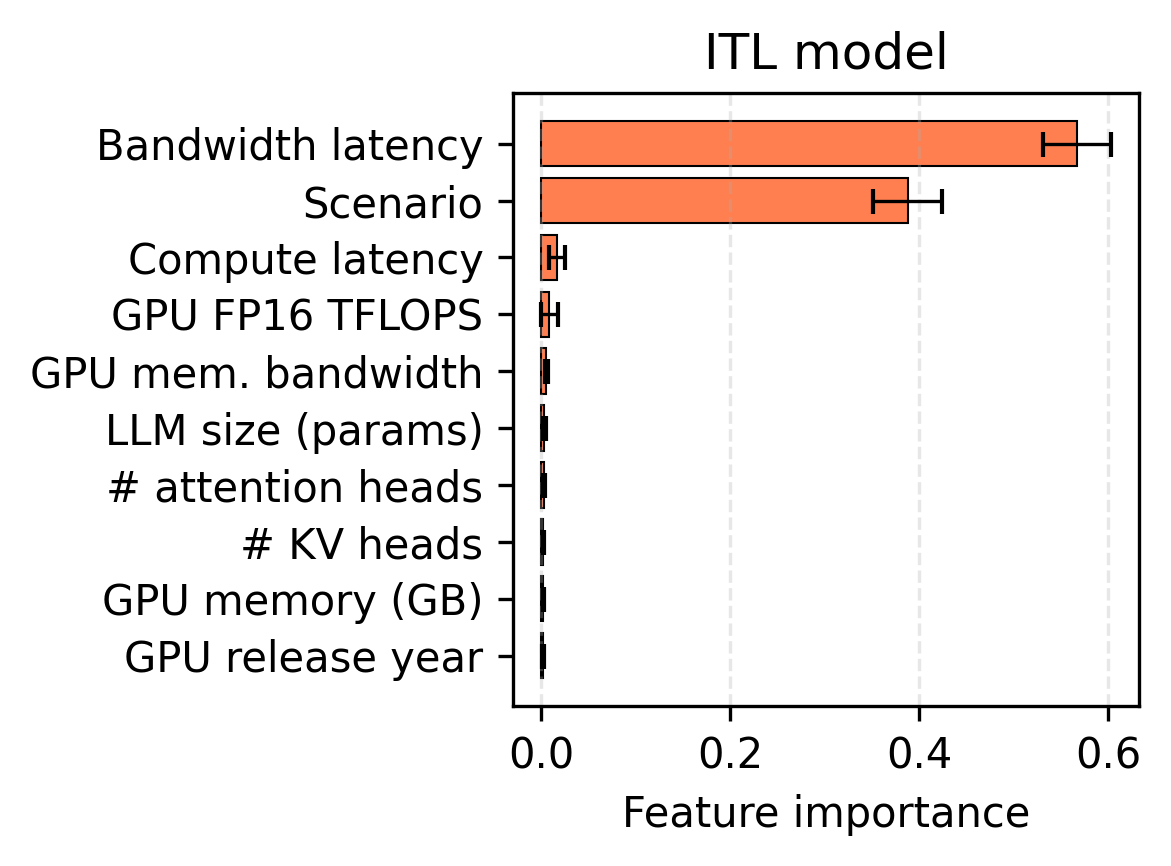}
    \end{subfigure}
    \caption{Feature importances of the trained XGBoost regressors for mean power draw and ITL. Both importances are averaged over all folds for LOGO validation. Error bars show $\pm1$ standard deviation across folds.}
    \label{fig:feature-importances}
\end{figure}

We also observe the feature importances of the trained XGBoost regressor to understand which are the main features used for mean power draw predictions in Figure~\ref{fig:feature-importances}. We see that scenario is the most important feature as expected, since the power draw pattern of offline and server scenarios is very different. This is followed by two memory-bandwidth-related features: GPU memory bandwidth and our engineered bandwidth latency. This is reasonable, as LLM inference is usually memory-bound and therefore the memory bandwidth and the model's size in memory determine the GPU utilization which influences its power draw. Moreover, we see that apart from scenario, GPU-based features have greater importance than LLM-based features, suggesting that the main differences in power draw are caused by the GPU and not the LLM used.

\subsection{ITL Prediction Results}
We evaluate the predictive model for ITL against a roofline-derived baseline that approximates ITL as the time to stream the LLM's FP16 weights from GPU memory once per token. This is calculated as $T_{ITL}=\frac{2P_{\mathrm{model}}\cdot 10^{9}}{(M_{bw} \cdot B)}$, where $P_{model}$ is the model size in parameters in billions, the factor 2 is bytes per parameter in FP16, $M_{bw}$ the GPU's memory bandwidth, and $B$ the expected batch size ($B=1$ for server, $B=256$ for offline deployments, corresponding to vLLM's \texttt{--max-num-seqs} default value). We utilize this baseline as ITL is mainly memory-bound in LLM inference. The baseline is analytical and therefore invariant to data splits. Results are shown in Table~\ref{tab:itl_results}.

\begin{table}[h]
\centering
\caption{Inter-token latency prediction performance for XGBoost (XGB) vs. a roofline-derived baseline. The baseline approximates ITL as the time to stream the LLM's FP16 weights from GPU memory once per token. The baseline is non-learning and therefore invariant to data splits.}
\label{tab:itl_results}
\scriptsize
\begin{tabular}{ccccccc}
\toprule
Model & Scenario & Strategy & MdAPE & Pearson $r$ & GPU $\tau$ & LLM $\tau$ \\
\midrule
XGB & Offline & CV & 12.8\% & 0.813 & 0.86 & 0.81 \\
XGB & Offline & LOGO & 24.9\% & 0.727 & 0.72 & 0.83 \\
XGB & Offline & LOLO & 15.6\% & 0.808 & 0.97 & 0.66 \\
Roofline & Offline & — & 80.2\% & 0.545 & 0.82 & 0.61 \\
\midrule
XGB & Server & CV & 4.9\% & 0.991 & 0.94 & 0.89 \\
XGB & Server & LOGO & 8.5\% & 0.972 & 0.78 & 0.88 \\
XGB & Server & LOLO & 5.6\% & 0.990 & 0.97 & 0.88 \\
Roofline & Server & — & 29.6\% & 0.991 & 0.88 & 0.91 \\
\bottomrule
\end{tabular}
\end{table}

The roofline baseline performs well in server scenarios for GPU and LLM rankings, confirming ITL is memory-bound. However, its absolute predictions achieve a MdAPE of 80.2\% in offline and 29.6\% in server scenarios, showing its limitations. Our approach significantly reduces these errors on CV ---i.e., for (GPU, LLM) unseen combinations---, achieving an MdAPE of 12.8\% and 4.9\% in offline and server scenarios respectively, with higher precision for rankings, shown by achieving a higher GPU Kendall $\tau$.
Generalization to unseen GPUs and LLMs is more challenging, increasing the MdAPE and decreasing our model's ranking capability.

The offline scenario is particularly challenging, as shown by the results of our model in LOGO (MdAPE=24.9\%, GPU $\tau$=0.72, LLM $\tau$=0.83) and LOLO (MdAPE=15.6\%, GPU $\tau$=0.97, LLM $\tau$=0.66). Notably, the roofline baseline matches or exceeds our model on GPU ranking under LOGO (both offline and server), while its MdAPE is at least 3$\times$ higher than our model's. This reflects a complementary relationship between analytical and learned models: roofline estimations have a high correlation with actual ITL and achieve a strong ranking performance, while our learned model captures the additive overheads (attention compute, kernel launch, scheduling), being capable of achieving higher accuracy in absolute ITL values than the baseline.

When looking at the feature importances learned by the ITL XGBoost regressor (Figure~\ref{fig:feature-importances}), we see that bandwidth latency is the most important feature, even more than scenario in this case, which again suggests that ITL is memory-bound in LLM inference. Scenario is second, and then compute latency followed by GPU FP16 TFLOPS, with significantly smaller importances. 
Incorporating the roofline estimate used as a baseline as an additional feature did not improve performance in this configuration, likely because the regressor can already reconstruct it from the existing feature set.

\begin{table}[ht]
\centering
\caption{LLMs evaluated in this study (42 dense models).}
\label{tab:llms}
\scriptsize
\begin{tabular}{llr}
\toprule
Model & Model Type & Params (B) \\
\midrule
openai-community/gpt2 & gpt2 & 0.124 \\
Qwen/Qwen2.5-0.5B-Instruct & qwen2 & 0.49 \\
nvidia/AceMath-1.5B-Instruct & qwen2 & 1.5 \\
Qwen/Qwen2.5-1.5B-Instruct & qwen2 & 1.54 \\
HuggingFaceTB/SmolLM2-1.7B-Instruct & llama & 1.7 \\
ibm-granite/granite-3.0-2b-instruct & granite & 2 \\
google/gemma-2-2b-it & gemma2 & 2.61 \\
microsoft/phi-2 & phi & 2.7 \\
tiiuae/Falcon3-3B-Instruct & llama & 3 \\
Qwen/Qwen2.5-3B-Instruct & qwen2 & 3.09 \\
meta-llama/Llama-3.2-3B-Instruct & llama & 3.21 \\
microsoft/Phi-3-mini-4k-instruct & phi3 & 3.8 \\
nvidia/Nemotron-Mini-4B-Instruct & nemotron & 4 \\
nvidia/Llama-3.1-Minitron-4B-Width-Base & llama & 5 \\
01-ai/Yi-1.5-6B-Chat & llama & 6 \\
EleutherAI/gpt-j-6b & gptj & 6.05 \\
mlabonne/NeuralBeagle14-7B & mistral & 7 \\
mlabonne/AlphaMonarch-7B & mistral & 7 \\
deepseek-ai/deepseek-llm-7b-chat & llama & 7 \\
berkeley-nest/Starling-LM-7B-alpha & mistral & 7 \\
nvidia/AceMath-7B-Instruct & qwen2 & 7 \\
allenai/OLMo-2-1124-7B-Instruct & olmo2 & 7 \\
internlm/internlm2-7b & internlm2 & 7 \\
mistralai/Mistral-7B-Instruct-v0.3 & mistral & 7.25 \\
Qwen/Qwen2.5-7B-Instruct & qwen2 & 7.62 \\
ibm-granite/granite-3.0-8b-instruct & granite & 8 \\
nvidia/Mistral-NeMo-Minitron-8B-Instruct & mistral & 8 \\
meta-llama/Llama-3.1-8B-Instruct & llama & 8.03 \\
01-ai/Yi-1.5-9B-Chat & llama & 9 \\
upstage/SOLAR-10.7B-Instruct-v1.0 & llama & 10.7 \\
google/gemma-3-12b-it & gemma3 & 12 \\
meta-llama/Llama-2-13b-chat-hf & llama & 13 \\
allenai/OLMo-2-1124-13B-Instruct & olmo2 & 14 \\
Qwen/Qwen3-14B & qwen3 & 14 \\
microsoft/Phi-3-medium-4k-instruct & phi3 & 14 \\
microsoft/phi-4 & phi3 & 14.7 \\
internlm/internlm2\_5-20b-chat & internlm2 & 20 \\
EleutherAI/gpt-neox-20b & gpt\_neox & 20 \\
upstage/solar-pro-preview-instruct & solar & 22.1 \\
mistralai/Mistral-Small-24B-Instruct-2501 & mistral & 24 \\
mistralai/Mistral-Small-Instruct-2409 & mistral & 24 \\
google/gemma-3-27b-it & gemma3 & 27 \\
\bottomrule
\end{tabular}
\end{table}
\section{Discussion \& Future Work}
\label{sec:discussion}

Evaluating both models across 42 dense LLMs and 8 server-grade GPUs reveals both the strengths and limits of predicting power and latency without target-system profiling. Our mean power draw model achieves high accuracy and is able to rank GPUs with good quality even on unseen GPUs, while our ITL model achieves precise absolute estimates and reliable GPU and LLM rankings under server-scenario workloads. Our models' predictions enable energy-aware deployment decisions such as GPU selection for a given LLM, model selection for a given GPU, capacity planning across heterogeneous hardware, and deployment energy estimates that contribute to reporting in line with emerging regulations such as the EU AI Act.

ITL prediction in offline scenarios remains harder than in server (LOGO MdAPE 24.9\% vs.\ 8.5\%). We theorize this is because the static features we currently use (i.e., LLM metadata and GPU specifications) cannot fully capture vLLM's continuous batching behavior. More workload-aware features, such as prompt and generation lengths or KV-cache usage, could narrow this gap. However, to maintain the generalization capacity of our model, these features need to be assumed by the operator or estimated from static data, which we leave for future work. Moreover, our evaluation focuses on dense models in single-GPU configurations; Mixture-of-Experts, quantized models, and multi-GPU deployments require separate modeling due to their distinct characteristics and remain future work. We also see promising directions in ensemble models that combine analytical and learned models as shown by Imai et al.~\cite{imai2024predicting}, exploiting the ranking performance of physical baselines such as roofline with the low absolute errors of learned models. 
\section{Conclusion}
\label{sec:conclusion}
As LLMs are increasingly deployed at scale, understanding and estimating the energy consumption of LLM inference deployments becomes essential for reducing energy usage and cost, assessing and reducing environmental impact, and complying with emerging regulatory frameworks. In this work, we introduced two predictive models ---to the best of our knowledge, the first to be evaluated on unseen GPUs--- to estimate mean GPU power draw and inter-token latency for LLM-GPU pairs using publicly available data, without requiring hardware access or profiling of the target pair. In our tests across 42 LLMs and 8 GPUs from different architectures, both models achieve median absolute percentage errors on unseen GPUs of $\leq3.4\%$ for offline and $\leq13.5\%$ for server scenarios for mean power draw and $\leq8.5\%$ for ITL 
in server scenarios. The hardest case for the ITL model is the offline scenario, where median absolute percentage error is $24.9\%$. Reducing this error requires workload-aware features or assumptions that we discuss as the main direction of future work in Section~\ref{sec:discussion}.

Our models are valuable for energy-aware placement decisions such as selecting GPUs for a target LLM, comparing LLMs for a given GPU, and supporting energy consumption and cost calculations in advance of deploying and profiling. Our models show their real value in cases in which commonly used heuristics such as TDP or TFLOPS per Watt are misleading. Figure~\ref{fig:gpu-spread} illustrates GPU selection for a low-load Llama 3.1 8B server with a 25ms ITL SLA: the L4's low 72W TDP suggests it as the one with lowest power consumption but our predictions show it violates the ITL SLA. Among compliant GPUs, the H100 has the highest TFLOPS per Watt at 2.09, but choosing it would draw 43\% more power than using an A30 (TDP: 165W, TFLOPS per Watt: 1.00). Since the server runs 24/7 regardless of GPU choice, this translates directly to a 43\% reduction in total GPU energy consumption. This decision contradicts TDP heuristics ---which would incorrectly select the L4 or the H100--- and would otherwise require benchmarking each configuration on physical hardware. Beyond energy savings, this reduces carbon emissions and operational costs, directly contributing to the sustainability and economic concerns that motivate our work.

Beyond improving accuracy on ITL prediction for offline scenarios, future work directions include extending our approach to support MoE and quantized models and integrating our models into energy-aware workload placement and request routing. In addition, as regulatory frameworks like the EU AI Act increasingly require energy reporting from AI deployments and this data is often not available with the granularity required from cloud operators, accurate predictive tools that operate on public data like ours offer a path toward accessible, transparent sustainability assessments and reporting.

\section*{Acknowledgment}
This work was partially supported by the European Engineering Learning Innovation and Science Alliance (EELISA).

\section*{Declaration on the Use of Generative AI}
The authors used Generative AI tools during the preparation of this manuscript for language editing purposes including grammar correction, spelling checks, and improvements to readability and clarity. The authors reviewed and verified all generated suggestions and take full responsibility for the content, analysis, results, and conclusions presented in this work.

\bibliography{sure26}

@inproceedings{theodorou2024energy,
  title={On Energy-aware and Verifiable Benchmarking of Big Data Processing targeting AI Pipelines},
  author={Theodorou, Georgios and Karagiorgou, Sophia and Kotronis, Christos},
  booktitle={2024 IEEE International Conference on Big Data (BigData)},
  pages={3788--3798},
  year={2024},
  organization={IEEE}
}

@article{argerich2024measuring,
  title={Measuring and improving the energy efficiency of large language models inference},
  author={Argerich, Mauricio Fadel and Pati{\~n}o-Mart{\'\i}nez, Marta},
  journal={IEEE Access},
  year={2024},
  publisher={IEEE}
}

@article{agrawal2024vidur,
  title={Vidur: A large-scale simulation framework for llm inference},
  author={Agrawal, Amey and Kedia, Nitin and Mohan, Jayashree and Panwar, Ashish and Kwatra, Nipun and Gulavani, Bhargav S and Ramjee, Ramachandran and Tumanov, Alexey},
  journal={Proceedings of Machine Learning and Systems},
  volume={6},
  pages={351--366},
  year={2024}
}

@inproceedings{tschand2025mlperf,
  title={MLPerf Power: Benchmarking the Energy Efficiency of Machine Learning Systems from $\mu$Watts to MWatts for Sustainable AI},
  author={Tschand, Arya and Rajan, Arun Tejusve Raghunath and Idgunji, Sachin and Ghosh, Anirban and Holleman, Jeremy and Kiraly, Csaba and Ambalkar, Pawan and Borkar, Ritika and Chukka, Ramesh and Cockrell, Trevor and others},
  booktitle={2025 IEEE International Symposium on High Performance Computer Architecture (HPCA)},
  pages={1201--1216},
  year={2025},
  organization={IEEE}
}

@inproceedings{strubell2020energy,
	title        = {Energy and policy considerations for modern deep learning research},
	author       = {Strubell, Emma and Ganesh, Ananya and McCallum, Andrew},
	year         = 2020,
	booktitle    = {Proceedings of the AAAI conference on artificial intelligence},
	number       = {09},
	pages        = {13693--13696}
}

@inproceedings{mlenergy-neuripsdb25,
    title={The {ML.ENERGY} Benchmark: Toward Automated Inference Energy Measurement and Optimization},
    author={Jae-Won Chung and Jeff J. Ma and Ruofan Wu and Jiachen Liu and Oh Jun Kweon and Yuxuan Xia and Zhiyu Wu and Mosharaf Chowdhury},
    year={2025},
    booktitle={NeurIPS Datasets and Benchmarks},
}

@inproceedings{samsi2023words,
  title={From words to watts: Benchmarking the energy costs of large language model inference},
  author={Samsi, Siddharth and Zhao, Dan and McDonald, Joseph and Li, Baolin and Michaleas, Adam and Jones, Michael and Bergeron, William and Kepner, Jeremy and Tiwari, Devesh and Gadepally, Vijay},
  booktitle={2023 IEEE High Performance Extreme Computing Conference (HPEC)},
  pages={1--9},
  year={2023},
  organization={IEEE}
}

@article{wilkins2024offline,
  title={Offline energy-optimal llm serving: Workload-based energy models for llm inference on heterogeneous systems},
  author={Wilkins, Grant and Keshav, Srinivasan and Mortier, Richard},
  journal={ACM SIGENERGY Energy Informatics Review},
  volume={4},
  number={5},
  pages={113--119},
  year={2024},
  publisher={ACM New York, NY, USA}
}

@inproceedings{kakolyris2025throttll,
  title={throttLL’eM: Predictive GPU Throttling for Energy Efficient LLM Inference Serving},
  author={Kakolyris, Andreas Kosmas and Masouros, Dimosthenis and Vavaroutsos, Petros and Xydis, Sotirios and Soudris, Dimitrios},
  booktitle={2025 IEEE International Symposium on High Performance Computer Architecture (HPCA)},
  pages={1363--1378},
  year={2025},
  organization={IEEE}
}

@techreport{burian2025_increasing_ai_energy,
  author       = {Burian, Vlad and Stalla‑Bourdillon, Arthur},
  title        = {The increasing energy demand of artificial intelligence and its impact on commodity prices},
  institution  = {European Central Bank},
  type         = {Economic Bulletin, {\textit{Focus} Box}},
  number       = {2/2025},
  year         = {2025},
  month        = may,
  note         = {Accessed: 2025‑07‑31},
  url          = {https://www.ecb.europa.eu/press/economic-bulletin/focus/2025/html/ecb.ebbox202502_03~8eba688e29.en.html}
}

@article{fu2025llmco2,
  title={Llmco2: Advancing accurate carbon footprint prediction for llm inferences},
  author={Fu, Zhenxiao and Chen, Fan and Zhou, Shan and Li, Haitong and Jiang, Lei},
  journal={ACM SIGENERGY Energy Informatics Review},
  volume={5},
  number={2},
  pages={63--68},
  year={2025},
  publisher={ACM New York, NY, USA}
}

@misc{aienergyscore-leaderboard,
    author = {Sasha Luccioni and Boris Gamazaychikov and Emma Strubell and Sara Hooker and Yacine Jernite and Margaret Mitchell and Scott Chamberlin},
    title = {AI Energy Score Leaderboard - December 2025},
    year = {2025},
    publisher = {Hugging Face},
    howpublished = "\url{https://huggingface.co/spaces/AIEnergyScore/Leaderboard}",
}

@inproceedings{niu2026tokenpowerbench,
  title={TokenPowerBench: Benchmarking the power consumption of LLM inference},
  author={Niu, Chenxu and Zhang, Wei and Li, Jie and Zhao, Yongjian and Wang, Tongyang and Wang, Xi and Chen, Yong},
  booktitle={Proceedings of the AAAI Conference on Artificial Intelligence},
  volume={40},
  number={38},
  pages={32582--32590},
  year={2026}
}

@article{patwari2025forecasting,
  title={Forecasting LLM inference performance via hardware-agnostic analytical modeling},
  author={Patwari, Rajeev and Sirasao, Ashish and Das, Devleena},
  journal={arXiv preprint arXiv:2508.00904},
  year={2025}
}

@article{ozcan2025quantifying,
  title={Quantifying the energy consumption and carbon emissions of llm inference via simulations},
  author={{\"O}zcan, Miray and Wiesner, Philipp and Wei{\ss}, Philipp and Kao, Odej},
  journal={arXiv preprint arXiv:2507.11417},
  year={2025}
}

@article{caravaca2025prompts,
  title={From Prompts to Power: Measuring the Energy Footprint of LLM Inference},
  author={Caravaca, Francisco and Cuevas, {\'A}ngel and Cuevas, Rub{\'e}n},
  journal={arXiv preprint arXiv:2511.05597},
  year={2025}
}

@techreport{iea2025energyai,
  author      = {{International Energy Agency}},
  title       = {Energy and {AI}},
  institution = {International Energy Agency},
  address     = {Paris},
  year        = {2025},
  month       = apr,
  url         = {https://www.iea.org/reports/energy-and-ai},
  note        = {License: CC BY 4.0}
}

@article{argerich2026watt,
  title={Watt Counts: Energy-Aware Benchmark for Sustainable LLM Inference on Heterogeneous GPU Architectures},
  author={Argerich, Mauricio Fadel and F{\"u}rst, Jonathan and Pati{\~n}o-Mart{\'\i}nez, Marta},
  journal={arXiv preprint arXiv:2604.09048},
  year={2026}
}

@article{rince2025ecologits,
  title={Ecologits: Evaluating the environmental impacts of generative AI},
  author={Rinc{\'e}, Samuel and Banse, Adrien},
  journal={Journal of Open Source Software},
  volume={10},
  number={111},
  pages={7471},
  year={2025}
}

@article{lei2026energy,
  title={The Energy Cost of Execution-Idle in GPU Clusters},
  author={Lei, Yiran and Fernandez, Jared and Kypriotis, Vasilis and Skarlatos, Dimitrios and Strubell, Emma and Sherry, Justine and Vosler, Daniel},
  journal={arXiv preprint arXiv:2604.04745},
  year={2026}
}

@article{krupp2026taking,
  title={This Is Taking Too Long-Investigating Time as a Proxy for Energy Consumption of LLMs},
  author={Krupp, Lars and Gei{\ss}ler, Daniel and Calatrava-Nicolas, Francisco M and Banwari, Vishal and Lukowicz, Paul and Karolus, Jakob},
  journal={arXiv preprint arXiv:2603.15699},
  year={2026}
}

@inproceedings{imai2024predicting,
  title={Predicting llm inference latency: A roofline-driven ml method},
  author={Imai, Saki and Nakazawa, Rina and Amaral, Marcelo and Choochotkaew, Sunyanan and Chiba, Tatsuhiro},
  booktitle={Annual Conference on Neural Information Processing Systems},
  year={2024}
}

\end{document}